# Synthesis and Characterization of Strontium Doped Barium Titanates using Neutron Diffraction Technique

I. B. Elius[1*], B. M. Asif[2], J. Maudood[1], T. K. Datta[1], A. K. M. Zakaria[3], S. Hossain[1], M. S. Aktar[1] and I. Kamal[3]

[1]*Institute of Nuclear Science and Technology, Bangladesh Atomic Energy Commission, GPO Box No. 3787, Dhaka-1000, Bangladesh*
[2]*Department of Nuclear Engineering, University of Dhaka, Dhaka, Bangladesh*
[3]*Bangladesh Atomic Energy Commission, Agargaon, Sher-e-Bangla Nagar, Dhaka-1207*

**Abstract**

In the present study, structural changes due to gradual doping of Sr in $BaTiO_3$ were investigated by both x-ray and neutron powder diffraction techniques. $Ba_{1-x}Sr_xTiO_3$ (x=0.0, 0.5 and 1.0) samples were synthesized by PVA evaporation method and purity was confirmed by x-ray diffraction experiment. After that, neutron diffraction experiments were carried out and diffraction data were analyzed by Rietveld least-square data refinement method using the computer program RIETAN-2000 and FullProf to determine various crystallographic structural parameters. The cation and anion position coordinates were also determined from the data refinement method which confirmed that Ba and Sr atoms possess tetrahedral A site and Ti atoms possess octahedral B site for all three samples $BaTiO_3$, $Ba_{0.5}Sr_{0.5}TiO_3$, and $SrTiO_3$, respectively. Moreover, the lattice parameter values indicate, $Ba_{1-x}Sr_xTiO_3$ structure undergoes a phase transition from tetragonal to cubic somewhere before x=0.5. The concurrences between observed and calculated profiles were excellent and well consistent with the previously reported values. Furthermore, MEM based analysis was done for charge density measurement.

*Keywords:* Neutron diffraction, Barium titanate, PVA evaporation method, X-ray diffraction, ferroelectric materials

## 1. Introduction

Barium titanate is one of the most studied ferroelectric ceramic materials which exhibit photoelectric and piezoelectric properties [1–3]. It has been widely used as electromechanical transducers, ceramic multilayer capacitors (CMC), microphones, in non-linear optics etc. [4]. Nanoparticles of Barium titanate are used as nano-carriers for drug delivery for its bio-compatibility, as capacitor energy storages in electric vehicles, in electro-optic modulation, in un-cooled sensors, in thermal capacitors etc. [3, 5]. Depending on the temperature, $BaTiO_3$ has four polymorphs (cubic, tetragonal, orthorhombic and rhombohedric consecutively). But in room temperature, it is tetragonal in structure [6]. When strontium is doped, $BaTiO_3$ goes through a phase transition from tetragonal to cubic [4, 7-8]. Because of its dielectric nonlinearity, large tenability and good dielectric thermal stability, strontium doped barium titanate materials have various uses as phase shifters, tunable RF-filters etc. [9-10].

Barium titanates are represented with a stoichiometric formula $ABO_3$ which stoichiometrically belongs to perovskite group of crystals. In this structure, A and B are cations and O is the anion (oxygen in this particular case). The ionic radii of A (typically $Ba^{2+}$, $Ca^{2+}$, $Sr^{2+}$, $La^{3+}$ etc.) is larger than B cations ($Ti^{3+}$, $Fe^{3+}$ etc.), where A cations reside at the corners of the structural cage having a 12-fold cuboctahedral coordination (0,0,0) and the B-ions are surrounded by the anions having a six-fold coordination at the very center (½, ½, ½) [3, 11-12].

The current study, $Ba_{1-x}Sr_xTiO_3$ (x = 0.0, 0.5, and 1.0) samples were synthesized by PVA (polyvinyl alcohol, $[CH2CH(OH)]_n$) evaporation method which is an environment-friendly, feasible, low cost and a low-temperature method [11, 13]. Finaly, both x-ray diffraction and neutron diffraction studies were performed for crystallographic studies via structural refinement.

## 2. Experimental

### 2.1 Sample preparation

The role of polyvinyl alcohol (average molecular weight 125000) in the current method is to make a homogeneous distribution of cations in its polymeric network structure and further, restrict the precipitation or segregation from the nitrate-PVA solution. Upon heating, PVA is decomposed and thus provides an oxidizing environment. $NO_2$ leaves the organic dark fluffy mixture during the evaporation process, which is the precursor. After the calcination process, porous-fine white powder is yielded as the organic parts evaporate in the form of $CO_2$, CO, water and other gas.

The perovskites $Ba_{1-x}Sr_xTiO_3$ (x = 0.0, 0.5, and 1.0) were prepared by the PVA evaporation method at the material synthesis laboratory in the Institute of Nuclear Science and Technology (INST) of Atomic Energy Research Establishment (AERE), Savar, Dhaka. All samples were prepared using High purity materials of $Ba(NO_3)_2$ (99% pure), $Sr(NO_3)_2$ (98% pure) and $TiO_2$ (99% pure) powders from Sigma Aldrich, India, in exact stoichiometric proportions. Then chemicals were weighted separately using a digital microbalance. At first, stoichiometric proportions of $Ba(NO_3)_2$, $Sr(NO_3)_2$ and $TiO_2$ were mixed with 200 ml water and the mixture was then heated in a magnetic heating stirrer to make it homogeneous. When the solutions started boiling, PVA of a proportional amount was mixed with the solutions. Different amounts of PVA was used for each sample.

Then the new PVA solutions were heated until the water was completely vaporized and the mixture turned into the ashes. After that, the ashes were heated in a muffle furnace in air environment at 300°C for 2 hours and then 500°C for 2 hours each (the temperature was raised and cooled down at a rate of 5°C/min). Then the mixtures of all the samples were ground in an agate mortar until the powder mixture became slippery. After grinding, all the samples were calcined in a muffle furnace at 700°C in air for 6.5 hours, in order to homogenize the end product. The raising and the cooling temperature was 5°C/min, respectively. Then the mixture of each sample was cooled down to room temperature in the furnace having air atmosphere at a rate of 5°C/min. Appropriate amount of polyvinyl alcohol (PVA) was added to each dried mixture as a binder and mixed intimately by agate mortar. It was then pressed into small pellets of cylindrical form with dimensions of 13mm diameter and width of 2-3 mm in a hydraulic press. The pellets did not require pre sintering as the ingredients do not have carbonates and higher oxides to be decomposed. The pellets were then sintered in air at 1250°C for 9 hours and subsequently cooled in the furnace at a slow rate. Finally, the pellets of each sample were pulverized into powder form again by grinding with the agate mortar. Time taken for each sample to convert from solid to powder was approximately 3 hours.

*Corresponding author: iftakhar.elius@gmail.com

*2.2 Data collection and refinement*

The samples were subjected to x-ray diffraction studies to confirm their phase purity. These studies were carried out using a Phillips PW 3040 *X'Pert Pro* x-ray diffractometer with Cu($K_\alpha$) radiation having an average wavelength of 1.54178Å, located at Atomic Energy Centre (AECD) of BAEC, Dhaka [14]. The XRD patterns were recorded within an angular range of $2\theta = 20°\sim70°$ having a step size of 0.02°. The comparison between the positions of the peaks in the XRD patterns of the samples and the standard samples confirmed the formation of desired phases [12, 15]. Moreover, the peak positions were indexed with *Chekcell* peak matching software. Absence of any other extra peaks indicated that no impurity or other chemicals were found in the samples. Space groups were identified using Chekcell software. The XRD patterns of the samples are presented in Fig. 1. Bragg's peaks at (100), (101), (111), (200), (210), (211), (220) and the splitting of (200) peak to afterwards (approaching to higher angle) suggested that the samples belong to tetragonal *(P4mm)* symmetry [16], which are shown in Fig. 1(a) and Fig. 1(b)). On the other hand, no splitting in any of the peaks in the SrTiO$_3$'s pattern indicated the formation of cubic symmetry *(Pm$\overline{3}$m)* [12, 17], it is shon in Fig. 1(c) .

Upon confirmation of the formation of a single-phase structure from x-ray diffraction data, long exposure neutron diffraction measurements were performed on the samples. X-ray photons interact with the outer shell electrons, whereas, neutrons interact with the nuclei of atoms and thus more specific and reliable when structural position data, bond lengths etc. are to measure. Each of the samples was subjected to neutron diffraction studies at Savar Neutron Diffractometer (SAND) facility located at the radial beam port-II of the TRIGA Mark-II research reactor (BTRR) of BAEC, AERE, Savar. The SAND facility comprises of an assembly of 15 He filled detectors which can maneuver around a sample table from $2\theta = 5°$ to 115°, with an angular resolution of 0.05°, covering an angular range of 110° [18]. The machine has The powdered samples were poured in a cylindrical vanadium can which was placed on a rotating table. The neutron beam emerging from the TRIGA Mark-II research reactor gets converted to monochrome using a Si-single crystal monochromator. Only the neutron beam having a wavelength of 1.5656Å can pass through the filtration system which is then collimated to impinge on the sample can. The scattering pattern recorded by Position Sensitive Detectors (PSD) is then processed and stored by the data acquisition system.

The recorded neutron diffraction data were analyzed by Rietveld least-square refinement method [19] using the computer program RIETAN-2000 [8] and FullProf [20] in order to find crystal structure. The Pseudo-Voigt function was used to model the peak profiles. The background was fitted using a six-factor polynomial. Parameters like scale factor, zero position parameter, isotropic profile parameters, lattice parameters, atomic positional parameters etc. were also refined. The refinements were done for the whole angular range, leaving no unmatched or unaccounted peaks. For barium titanate sample, split Pseudo-Voigt function was used to model the peaks; other refinements were carried out in a similar manner.

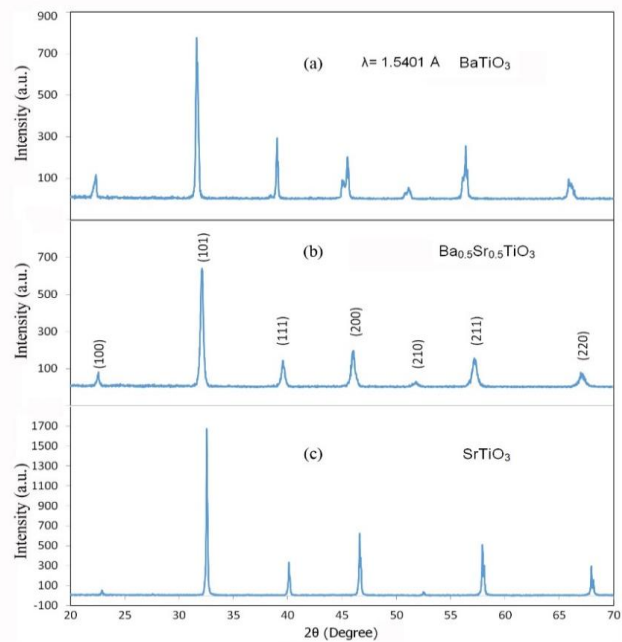

**Fig. 1:** X-ray diffraction pattern of (a) BaTiO$_3$, (b) Ba$_{0.5}$Sr$_{0.5}$TiO$_3$ and (c) SrTiO$_3$, respectively

## 3. Results and Discussion

It is well evident from previous studies that BaTiO$_3$ is tetragonal in room temperature and assumes the cubic, *Pm$\overline{3}$m* phase at 123°C (396K) [21]. A. Kaur et al. observed that this temperature can be reduced when strontium is doped gradually in barium titanate systems [21]. In the barium titanate pattern, the presence of splitting in *(h00)* type peaks is a clear indication of the tetragonal (P4mm) phase, which is also consistent with x-ray data. Later, these NUCLEAR SCIENCE AND APPLICATIONS                                   Vol. 28. No. 1&2  2019bifurcated peaks were seen to coalesce with each other upon Sr doping. The diffraction data of $Ba_{0.5}Sr_{0.5}TiO_3$ were refined with both of the symmetries, as it had a c/a ratio closer to unity, thus it is pseudo-cubic. All observed symmetry and crystal system of the samples are given in Table 1.

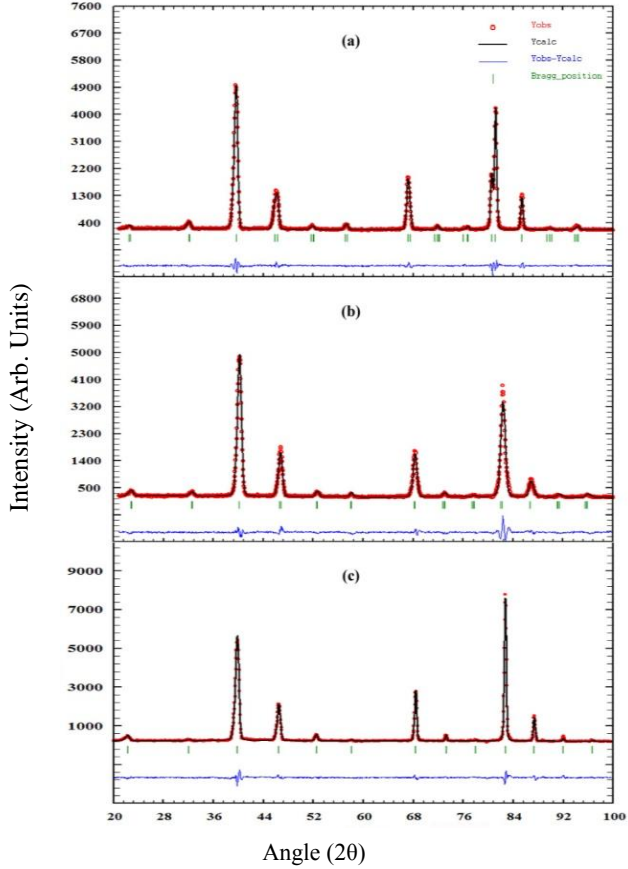

**Fig. 2:** Neutron diffraction patterns of (a) $BaTiO_3$, (b) $Ba_{0.5}Sr_{0.5}TiO_3$ and (c) $SrTiO_3$ refined using FullProf software where the red dots represent the experimental data, the black line is the simulated pattern and the blue line below is the difference curve. The peak splitting at $2\theta \sim 80°$ indicates the tetragonal phase of $BaTiO_3$.

**Table 1:** Symmetry and Crystal System of the samples

| Sample | Symmetry | Crystal System |
| --- | --- | --- |
| $BaTiO_3$ | P4mm | Tetragonal |
| $Ba_{0.5}Sr_{0.5}TiO_3$ | $Pm\bar{3}m$ | Pseudo-Cubic |
| $SrTiO_3$ | $Pm\bar{3}m$ | Cubic |

The structural refinements using FullProf refinement software are presented in Fig. 2. It can be seen that the inferred model (the dark solid line, generated from the initial crystallographic parameters which were then further refined) conforms agreeably with the diffraction data (the red data points). As a result the difference curve ($Y_{obs} \sim Y_{calc}$) is sufficiently suppressed too.

It is evident that various measured structural parameters of the series $Ba_{1-x}Sr_xTiO_3$ are reasonably agreeable to the literature reference values [3, 22-23]. All these indicate an excellent synthesis of the samples and characterization by the neutron diffraction technique. The reliability factors like profile factor ($R_p$), Weighted factor ($R_w$), Expected Weighted Profile Factor ($R_e$), Structure Factor ($R_f$), Braggs R-factor ($R_{Bragg}$) and Goodness of Fit or Chi-square value ($\chi^2$) obtained from the best fitting are given in Table 2. The patterns and reliability factors indicate that there is an excellent agreement between observed and calculated patterns.

**Table 2:** The reliability factors of $BaTiO_3$, $Ba_{0.5}Sr_{0.5}TiO_3$ and $SrTiO_3$

| R-factors | $BaTiO_3$ | $Ba_{0.5}Sr_{0.5}TiO_3$ | $SrTiO_3$ |
| --- | --- | --- | --- |
| $R_p$ | *11.900* | *12.50* | *13.00* |
| $R_w$ | *11.100* | *11.90* | *12.00* |
| $R_e$ | *9.660* | *9.350* | *9.240* |
| $R_f$ | *3.700* | *3.170* | *3.850* |
| $R_{Bragg}$ | *2.740* | *2.290* | *2.260* |
| $\chi^2$ | *1.320* | *1.621* | *1.686* |

Formula used to determine various refinement parameters are given below.

Profile factor, $R_P = \left[ \dfrac{\sum_i |Y_i^{obs} - Y_i^{calc}|}{\sum_i Y_i^{obs}} \right]$ (1)

Weighted factor, $R_W = \left[ \dfrac{\sum_i w_i |Y_i^{obs} - Y_i^{calc}|^2}{\sum_i w_i Y_i^{obs^2}} \right]^{\frac{1}{2}}$ (2)

Expected weighted profile factor, $R_E = \left[ \dfrac{N-P+C}{\sum_i w_i Y_i^{obs^2}} \right]^{\frac{1}{2}}$ (3)

Structure factor, $R_f = \dfrac{\sum_{hkl} ||F_{hkl}^{in}|^2 - |F_{hkl}^{out}|^2|}{\sum_{hkl} |F_{hkl}^{in}|^2}$ (4)

Bragg factor, $R_{Bragg} \left[ \dfrac{\sum_k |I_k^{obs} - I_k^{calc}|}{\sum_k I_k^{obs}} \right]$ (5)

Chi-squared factor, $\chi^2 = \left( \dfrac{R_{wP}}{R_E} \right)^2$ (6)

The bond lengths were measured with *BondStr* computer code which is based on *CrysFML* (Crystallographic Fortran 95 Modules Library) [24]. The calculated and expected bond lengths are given in Table 3. The program utilizes the standard structural database to calculate an expected bond-valence distance and compare it with distances calculated from structural positions and atomic radii. The Ba-O length increases from *2.8286* Å to *2.8345* Å when Sr is doped in $BaTiO_3$. Sr-O bond length is decreased from 2.7953 Å to 2.7542 Å when Ba is entirely replaced by Sr. On the other hand, the bond length of Ti-O is gradually reduced from

59

2.0344 Å to 1.9771 Å when 50% Ba is replaced by Sr and from 1.9771 Å to 1.947 Å when Sr completely replaced Ba. From data refinement, various structural parameters were determined. Structural parameters and goodness of fit values obtained from both FullProf and Rietan-2000 are given below in Table 4. The density of the compound as calculated from neutron diffraction reduces from 6.050 g/cm$^3$ as the atomic mass of Ba atom is 137.33 amu, which is being replaced by much lighter Sr (87.62 amu) atoms. The Sr atoms are also substantially smaller than Ba atoms which accounts for the reduction of cell volume when Sr is added to $BaTiO_3$, shown in Table 5. The oxygen position parameter is close to the ideal value for FCC, which is 0.500. As Sr is replaced by Ba, the oxygen atoms have moved further from the ideal position. The half-width parameters (i.e. U, V & W) are also given in Table 5. As the samples are bulk (of μm scale), no direct relationship could be calculated from these values as empirical formulas like the Scherrer's formula does not hold in such scale.

**Table 3:** Bond lengths of (Ba-O), (Sr-O) and (Ti-O) in $BaTiO_3$, $Ba_{0.5}Sr_{0.5}TiO_3$ and $SrTiO_3$ are given below with their expected values

| Bond | $BaTiO_3$ | | $Ba_{0.5}Sr_{0.5}TiO_3$ | | $SrTiO_3$ | |
|---|---|---|---|---|---|---|
| | Calculated (Å) | Expected (Å) | Calculated (Å) | Expected (Å) | Calculated (Å) | Expected (Å) |
| Ba-O | 2.8286 | 2.8256 | 2.8345 | - | - | - |
| Sr-O | - | - | 2.7953 | 2.793(2) | 2.7542 | 2.7412 |
| Ti-O | 2.0344 | 2.0315 | 1.9771 | 1.974(9) | 1.947(7) | 1.9475 |

**Table 4:** Comparison of structural parameters and goodness of fit yielded by two software FullProf and Rietan-2000.

| Sample | a (Å) | | b (Å) | | c (Å) | | $\chi^2$ | | Structure |
|---|---|---|---|---|---|---|---|---|---|
| | Fullprof | Rietan | Fullprof | Rietan | Fullprof | Rietan | Fullprof | Rietan | |
| $BaTiO_3$ | 3.9884 | 3.9957 | 3.9884 | 3.9957 | 4.0287 | 4.0308 | 1.320 | 1.402 | Tetragonal |
| $Ba_{0.5}Sr_{0.5}TiO_3$ | 3.9495 | 3.9517 | 3.9495 | 3.9517 | 3.9580 | 3.9518 | 1.621 | 1.705 | Pseudo-Cubic |
| $SrTiO_3$ | 3.8950 | 3.9039 | 3.8950 | 3.9039 | 3.8950 | 3.9034 | 1.686 | 1.746 | Cubic |

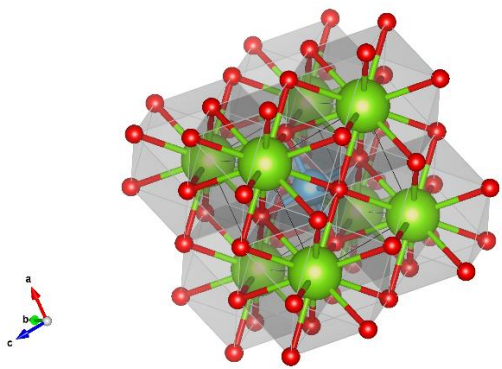

**Fig. 3:** Polyhedral view of $BaTiO_3$. The central blue sphere is the $Ti^{4+}$ ion, the green spheres are $Ba^{2+}/Sr^{2+}$ ions and the red spheres are oxygen anions. (Image generated by *VESTA*)

The polyhedral view is generated by software package VESTA (Fig. 3) using the crystallographic information file yielded from the refinement. This particular one is actually a unit cell of $BaTiO_3$ molecule, the planes enclosed by the bonds construct the polyhedra. The structural parameters of $BaTiO_3$ are in agreement with the studies of *C. J. Xiao et al.* [25]. Similarly, the parameters of $SrTiO_3$ are also in good agreement with *Valeri Petkov et al.* [26] and *Wontae Chang et al.* [27]. *Valeri Petkov et al.* investigated the structural properties of $Ba_{0.5}Sr_{0.5}TiO_3$, using Synchrotron radiation scattering studies and atomic pair distribution function (PDF) techniques (the refined and calculated results were 3.985(4) Å and 3.979(6) Å) [26-27].

**Table 5:** Change of cell volume, density of the compounds, Oxygen position parameter and half-width parameters (U, V, W) of the compounds.

| Parameters | $BaTiO_3$ | $Ba_{0.5}Sr_{0.5}TiO_3$ | $SrTiO_3$ |
|---|---|---|---|
| density (volumic mass) of the compound (g/cm$^3$) | 6.050 | 5.604 | 5.156 |
| Cell Volume (Å$^3$) | 64.0067 | 61.7387 | 59.0918 |
| Oxygen Position Parameters | (x,y,z)= (0.0, 0.0, 0.47877) | (x,y,z)= (0.0, 0.0, 0.48635) | (x,y,z)= (0.0, 0.0, 0.49458) |
| H.W. Parameters | U= 1.06235 V= -1.81614 W= 0.98834 | U= 3.05057 V= -2.79026 W= 1.20109 | U= 1.260209 V= -2.122915 W= 1.076660 |

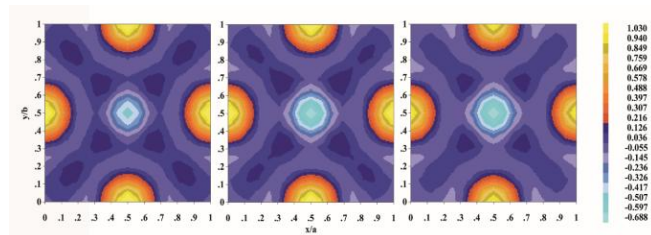

**Fig. 4:** The charge density maps of $BaTiO_3$, $Ba_{0.5}Sr_{0.5}TiO_3$ and $SrTiO_3$ (from left to right) along z/c=0.5 plane, the charge density in charges/Å$^3$.

The interatomic charge density and bonding characteristics were measured from the diffraction data by Maximum Entropy Method (MEM) based technique using GFOURIER software. Figure 4 exhibits the charge densities of the $BaTiO_3$, $Ba_{0.5}Sr_{0.5}TiO_3$ and $SrTiO_3$ along a plane *z/c=0.5* parallel to *(001)* plane. The changes in the charge densities may be attributed to the changes in the oxygen anions as the phase transition occurs and x/a ratio decreases. The program calculates the charge or scattering density $\rho(r)$ of a unit cell [28-29]. Where,

$$\rho(r) = \frac{1}{V} \sum_{H} F(H) exp\{-2\pi i(H \cdot r)\} \quad (7)$$

Here, V is the volume of a unit cell, **H** is a vector which represents reciprocal lattice, **r** is the position vector, F(**H**) is a function of **H** which are coefficients of the complex fourier transform. (The density yielded can be represented by number of charges/Å$^3$).

In Fig. 4, the central blue contour (and area inside) is the $Ti^{4+}$ ion and the four surrounding atoms at the middle of each side of the cube/pseudo-cube are the $O^{2-}$ anions. The charge of the scale is inverted, as it considers the electrons charge as unit. As Sr is doped, gradual increment of the central blue area indicates the reduced electron density in the central region, which can be attributed to reduced electron affinity of Sr (0.052 eV) compared to that of Ba (0.14462 eV) [30-31].

## 4. Conclusion

In summary, $BaTiO_3$, $Ba_{0.5}Sr_{0.5}TiO_3$ and $SrTiO_3$ were synthesized via feasible PVA evaporation method. The samples were subjected to XRD and neutron diffraction studies. The Neutron diffraction data were refined using Rietan-2000 and FullProf. All the refinements had an excellent goodness of fit value which indicated the accuracy of the refinement technique. During the refinement method, the profile parameters were varied one by one until the observed and simulated pattern totally matched. All the structure factors were in agreement with the established values of the previous studies. The half-width parameters, anion position parameters, cell volume, density etc. were measured using the *FullProf* code.

It was observed that, when 50% of the Ba atoms were replaced by the Sr atoms, the structure changed from tetragonal to cubic. This change may occur in the sample $Ba_{1-x}Sr_xTiO_3$ even for x<0.5. Therefore, investigation of the exact composition at which the phase change occurs, requires a smaller increment of Sr dopant and study of their x-ray and neutron diffraction analyses. The charge density maps, generated using Maximum Entropy Method (MEM) exhibited the reorientation of the density contours due to elongation of the cell along a particular direction.

## Acknowledgement

The authors are thankful to the personnel of the Center for Research Reactor (CRR), Bangladesh Atomic Energy Commission (BAEC) for the operation of the reactor during the Neutron Diffraction studies. The technical personnel of Neutron Diffraction group, RNPD are gratefully acknowledged for their kind help during sample preparation and analysis for this study.